\documentclass[12pt,reqno]{article}
\usepackage{amsfonts}
\usepackage{amsmath}
\usepackage{amsbsy}

\usepackage{amssymb,latexsym}
\numberwithin{equation}{section}

\begin{document}
 \allowdisplaybreaks[1]
\title{Algebraic Integration of Sigma Model Field Equations}
\author{Nejat T. Y$\i$lmaz\\
Department of Mathematics
and Computer Science,\\
\c{C}ankaya University,\\
\"{O}\u{g}retmenler Cad. No:14,\quad  06530,\\
 Balgat, Ankara, Turkey.\\
          \texttt{ntyilmaz@cankaya.edu.tr}}
\maketitle
\begin{abstract}
We prove that the dualization algebra of the symmetric space coset
sigma model is a Lie algebra and we show that it generates an
appropriate adjoint representation which enables the local
integration of the field equations yielding the first-order ones.
\end{abstract}

\section{Introduction}
The field content of the symmetric space coset sigma model
consists of scalar fields that parametrize the target manifold
which is a homogeneous and a Riemannian globally symmetric space.
By doubling the field content via the introduction of higher order
dual fields one can realize the theory through the construction of
an enlarged coset. The origin of this method lies in the
dualization of supergravity theories \cite{julia1,julia2} whose
scalar sectors correspond to above mentioned type of sigma models.
The most important element of this enlarged realization of
mentioned theories is the construction of the dualized coset
parametrizing algebra which for the case of the pure sigma model
is a deformation of the original coset algebra. Although the
geometrical construction of this extended formulation is not yet
well known the dualized algebra for a general coset sigma model
with a symmetric target space is derived in \cite{nej1,nej2}. In
these works the first-order field equations of the theory are also
obtained as consistency conditions embedded in the method of
dualization. However either of the works lack the direct algebraic
connection of these first-order equations with the second-order
ones which arise from the least action principle.

In this work we present a rigorous proof which shows that the
dualized coset algebra obtained in \cite{nej1,nej2} is indeed a
Lie algebra. The main perspective of our proof will be to show
that in the most general terms (for an arbitrary sigma model) the
commutation relations of the dualized algebra which comes out to
be a deformation of the ordinary coset algebra satisfies the
Jacobi identities. Following this we will also discuss that being
a Lie algebra the dualized coset algebra admits a natural adjoint
representation for the original coset algebra which is a Lie
subalgebra of the former. Finally we show that when one assumes
this natural adjoint representation generated by the dualized
algebra one can locally relate the first-order equations derived
in \cite{nej1,nej2} to the second-order field equations
algebraically. Namely starting from the first-order equation
anzats by applying an exterior derivative we will show that one
obtains the second-order field equations under the special
representation generated by the dualized coset algebra.

Section two which is a rather formal one inspects all the possible
conditions of a generic dualized coset algebra thus it presents a
complete proof of the Lie algebra structure of it for an arbitrary
coset sigma model. Section three discusses the natural adjoint
representation of the original coset algebra suggested within this
scheme. The last section proves that the first-order equations
which appear in \cite{nej1,nej2} are indeed the true local ones
which can be obtained from the second-order field equations of the
theory by locally abolishing an exterior derivative when one
chooses the above-mentioned particular representation.
\section{The Dualized Coset Algebra}
The dualized coset algebra of a generic symmetric space sigma
model is derived in \cite{nej1} and \cite{nej2}. It is generated
by the set of generators
\begin{equation}\label{eq1}
 \{H_{i},E_{\alpha},\widetilde{H}_{i},\widetilde{E}_{\alpha}\},
\end{equation}
where the first two set of generators correspond to a subset of
the Cartan-Weyl basis of the global symmetry group of the sigma
model Lagrangian which generate the solvable Lie subalgebra $s$.
Here for $i=1,\cdots,r$ the generators $H_{i}$ form a subset of
the Cartan generators of the Lie algebra of the global symmetry
group of the sigma model and $E_{\alpha}$ generate the root
subspaces of the non-compact positive roots $\Delta_{nc}^{+}$
\cite{nej1,nej2}. The last two set of generators are the duals of
the former. The commutation relations of the dualized coset
algebra that is generated by \eqref{eq1} can be given as
\begin{subequations}\label{eq2}
\begin{gather}
[H_{i},H_{j}]=[H_{i},\widetilde{H}_{j}]=[E_{\alpha},\widetilde{H}_{j}]=[\widetilde{E}_{\alpha},\widetilde{H}_{j}]
=[\widetilde{H}_{i},\widetilde{H}_{j}]=[\widetilde{E}_{\alpha},\widetilde{E}_{\beta}]=0,\notag\\
\notag\\
[H_{i},E_{\alpha}]=\alpha_{i}E_{\alpha}\quad,\quad[E_{\alpha
},E_{\beta }]=N_{\alpha ,\beta }E_{\alpha +\beta }
\quad\text{if}\quad
\alpha +\beta\in \Delta,\notag\\
\notag\\
 [E_{\alpha },E_{\beta
}]=0\quad\text{if} \quad\alpha +\beta\notin
\Delta,\notag\\
\notag\\
[E_{\alpha },\widetilde{E}_{\alpha }]=\frac{1}{4}\overset{r}{\underset{j=1}{%
\sum }}\alpha _{j}\widetilde{H}_{j}\quad,\quad
[H_{j},\widetilde{E}_{\alpha }]=-\alpha _{j}\widetilde{E}_{\alpha
},\notag\\
\notag\\
 [E_{\alpha },\widetilde{E}_{\beta
}]=0\quad\text{if}\quad\alpha
-\beta\notin\Delta,\notag\\
\notag\\
\text{or}\quad\alpha
-\beta\in\Delta\quad\text{but}\quad\beta-\alpha\notin\Delta_{nc}^{+},\notag\\
\notag\\
[E_{\alpha },\widetilde{E}_{\beta }]=N_{\alpha ,-\beta }\widetilde{E}%
_{\gamma }\quad\text{if}\quad\alpha
-\beta\in\Delta,\notag\\
\notag\\
\quad\beta-\alpha\in\Delta_{nc}^{+},\quad\text{and}\quad\alpha
-\beta =-\gamma.\tag{\ref{eq2}}
\end{gather}
\end{subequations}
Here $\Delta$ corresponds to the roots of the Lie algebra of the
global symmetry group of the sigma model. $\alpha_{i}$ are the
root vector components and the real coefficients $N_{\alpha\beta}$
are the structure constants corresponding to the commutation
relations of the root subspace generators $E_{\alpha}$. We should
remark that if $\alpha$ and $\beta$ are noncompact positive roots
and if $\alpha+\beta\in \Delta$ then $\alpha+\beta$ must also be a
noncompact positive root since if it is not then
\begin{equation}\label{eq3}
[E_{\alpha },E_{\beta }]=N_{\alpha ,\beta }E_{\alpha+\beta}\notin
s,
\end{equation}
which causes a contradiction for the closure of the solvable Lie
subalgebra. Now we will introduce the notation
\begin{equation}\label{eq4}
\{T_{m}\}\equiv\{H_{i},E_{\alpha}\}\quad,\quad\widetilde{T}_{m}
\equiv\{\widetilde{H}_{i},\widetilde{E}_{\alpha}\},
\end{equation}
so that the index $m$ is split into two sets
\[
m=\overbrace{1,\cdots\cdots,r}^{i,j,k,\cdots},\underbrace{r+1,r+2,\cdots\cdots,\text{dim}s}
_{\alpha,\beta,\gamma,\cdots}.
\]
In other words
\begin{subequations}\label{eq5}
\begin{gather}
T_{1}=H_{1}\quad,\quad T_{2}=H_{2}\quad,\quad
\cdots\cdots\quad,\quad
T_{r}=H_{r},\notag\\
\notag\\
\widetilde{T}_{1}=\widetilde{H}_{1}\quad,\quad\widetilde{T}_{2}=\widetilde{H}_{2}\quad,\quad
\cdots\cdots\quad,\quad\widetilde{T}_{r}=\widetilde{H}_{r},\notag\\
\notag\\ T_{r+1}=E_{\alpha}\quad,\quad
T_{r+2}=E_{\beta}\quad,\quad\cdots\cdots,\notag\\
\notag\\
\widetilde{T}_{r+1}=\widetilde{E}_{\alpha}\quad,\quad\widetilde{T}_{r+2}=\widetilde{E}_{\beta}\quad,\quad\cdots\cdots.
\tag{\ref{eq5}}
\end{gather}
\end{subequations}

The commutation relations in \eqref{eq2} can more compactly be
written as
\begin{subequations}\label{eq6}
\begin{gather}
[H_{i},H_{j}]=0\quad,\quad[H_{i},E_{\alpha}]=\alpha_{i}E_{\alpha}\quad,\quad[E_{\alpha
},E_{\beta }]=0\quad\text{if} \quad\alpha +\beta\notin
\Delta,\notag\\
\notag\\
[E_{\alpha },E_{\beta }]=N_{\alpha ,\beta }E_{\alpha +\beta }
\quad\text{if}\quad
\alpha +\beta\in \Delta,\notag\\
\notag\\
[\widetilde{T}_{m},\widetilde{T}_{n}]=0\quad,\quad[E_{\gamma
},\widetilde{T}_{m}]=\widetilde{f}_{\gamma
m}^{n}\widetilde{T}_{n}\quad,\quad[H_{j
},\widetilde{T}_{m}]=\widetilde{g}_{j
m}^{n}\widetilde{T}_{n},\tag{\ref{eq6}}
\end{gather}
\end{subequations}
where the structure constant matrices $\widetilde{f}_{\gamma}$ and
$\widetilde{g}_{j}$ in partitioned form can be given as

\begin{equation}\label{eq7}
\widetilde{f}_{\gamma}=\overset{\scriptstyle i\scriptstyle
=\scriptstyle1,\scriptstyle2,\scriptstyle\cdots\scriptstyle\cdots\scriptstyle\cdot\scriptstyle
r\quad\quad\quad\scriptstyle\alpha,\scriptstyle\beta,\scriptstyle\cdots
\scriptstyle\gamma,\scriptstyle\cdots\scriptstyle\cdot\scriptstyle\cdot\scriptstyle\cdot}
{\left(\begin{array}{c|c} \begin{pmatrix} &&&&\\
&&\text{\Large{0}}&&\\&&&&
\end{pmatrix}&
\begin{pmatrix} & \:\:\frac{\gamma_{1}}{4}
\\\text{\Large{0}} &\:\:\vdots&\text{\Large{0}}\\&\:\:\frac{\gamma_{r}}{4} \end{pmatrix}    \\
\hline
\begin{pmatrix}&&&&\\&&\text{\Large{0}}&&\\&&&&\end{pmatrix}&
\begin{pmatrix}&&&&\\&&\text{\Large{$R$}}_{\gamma}&&\\&&&&\end{pmatrix}
\end{array}\right)}\begin{array}{c}\overset{\overset{\underset{\scriptstyle\shortparallel}{\scriptstyle
i}}{ \scriptstyle 1}}{\underset{\scriptstyle
r}{\overset{\scriptstyle\cdot}{\overset{\scriptstyle\cdot}{\overset{\scriptstyle\cdot}{\scriptstyle\cdot}
}}}}\\\scriptstyle\alpha\\\scriptstyle
\beta\\\scriptstyle\vdots\\\\\end{array},
\end{equation}
\begin{equation}\label{eq8}
\widetilde{g}_{j}=-\overset{\scriptstyle i\scriptstyle
=\scriptstyle1,\scriptstyle2,\scriptstyle\cdots\scriptstyle\cdots\scriptstyle\cdot\scriptstyle
r\quad\quad\quad\scriptstyle\alpha,\:\:\:\scriptstyle\beta,\scriptstyle\cdots
\scriptstyle\cdots\scriptstyle\cdot\scriptstyle\cdot\scriptstyle\cdot}
{\left(\begin{array}{c|c} \begin{pmatrix} &&&&\\
&&\text{\Large{0}}&&\\&&&&
\end{pmatrix}&
\begin{pmatrix} &&&&
\\&&\text{\Large{0}} &&\\&&&&\end{pmatrix}    \\
\hline
\begin{pmatrix}&&&&\\&&\text{\Large{0}}&&\\&&&&\end{pmatrix}&
\begin{pmatrix}\alpha_{j}&&\text{\Large{0}}\\&\beta_{j}&\\\text{\Large{0}}&&\ddots\end{pmatrix}
\end{array}\right)}\begin{array}{c}\overset{\overset{\underset{\scriptstyle\shortparallel}{\scriptstyle i}}{
\scriptstyle 1}}{\underset{\scriptstyle
r}{\overset{\scriptstyle\cdot}{\overset{\scriptstyle\cdot}{\scriptstyle\cdot
}}}}\\\underset{\scriptstyle\alpha}{\mathstrut}\\\scriptstyle
\beta\\\scriptstyle\vdots\\\\\end{array},
\end{equation}
where we have defined
\begin{equation}\label{eq9}
\text{\large{$R$}}_{\gamma}= \left(\begin{matrix}
\overset{\scriptstyle\alpha}{\mathstrut
0}&\overset{\:\:\scriptstyle\beta}{\mathstrut}&
\overset{\:\:\scriptstyle\cdots}{\mathstrut}&\overset{\scriptstyle\kappa}{\mathstrut
0}&\overset{\scriptstyle\cdots}{\mathstrut}
&\overset{\scriptstyle\cdots}{\mathstrut}&\overset{\scriptstyle\cdots}{\mathstrut}
\\
&\:\:0&&\vdots&&\\&&\:\:\ddots&\vdots&&\\\\&&&0&&\\0&\cdots&0&(\tau,\kappa:=\gamma)&0&\cdots&0\\&&&0&\ddots&\\&&&\vdots&&0&
\\&&&0&&&0\end{matrix}\right)
\begin{array}{c}\overset{\scriptstyle\alpha}{\mathstrut}\\\overset{\scriptstyle\beta}{\mathstrut}\\\scriptstyle\vdots\\\scriptstyle\vdots
\\\overset{\scriptstyle\tau}{\mathstrut}\\
\scriptstyle\vdots\\\scriptstyle\vdots\end{array}.
\end{equation}
Here the non-zero entry in a certain column and a row is defined
as
\begin{equation}\label{eq10}
(\tau,\kappa:=\gamma)=N_{\gamma,-\kappa}\quad\text{if}\quad\gamma-\kappa=-\tau,\quad\text{otherwise}
\quad(\tau,\kappa:=\gamma)=0.
\end{equation}
In general
\begin{subequations}\label{eq11}
\begin{gather}
(\text{\large{$R$}}_{\gamma})^{\alpha}_{\:\:\:\beta}=N_{\gamma,-\beta}\quad\text{if}\quad\gamma-\beta=-\alpha,\notag\\
\notag\\
(\text{\large{$R$}}_{\gamma})^{\alpha}_{\:\:\:\beta}=0\quad\text{if}\quad\gamma-\beta\neq-\alpha.\tag{\ref{eq11}}
\end{gather}
\end{subequations}
Before going further we will state some facts about the matrix
$R_{\alpha}$;
\begin{itemize}
\item if for a fixed column $\beta$, $\alpha -\beta\notin\Delta,$
or $\alpha -\beta\in\Delta$ but for none of the rows $ \gamma$, $
\alpha -\beta\neq-\gamma$ then the column $\beta$ has all null
entries,
 \item if again for a fixed column $\beta$, $\alpha
-\beta\in\Delta,$ and if $\alpha -\beta=-\gamma$ and $\alpha
-\beta=-\tau$ $\Rightarrow$ $\gamma=\tau$, thus at a column
$\beta$ if all the entries are not null then there is a single
non-zero entry which is $N_{\alpha,-\beta}$,

\item if for a fixed row $\gamma$, $\nexists$ a column $\beta$
such that $\alpha -\beta=-\gamma$ then all the entries in $\gamma$
are null. In other words if $\alpha+\gamma\notin\Delta$ then
$\nexists$ $\beta$ such that $\alpha -\beta=-\gamma$ and the
$\gamma$ row is null. However if $\alpha+\gamma\in\Delta$ then
from our previous discussion $\alpha+\gamma\in\Delta_{nc}^{+}$ so
$\exists$ a column $\beta$ such that $\alpha -\beta=-\gamma$ and
the row $\gamma$ has a non-zero entry,

 \item if for a fixed row $\gamma$, $\exists$ two columns
$\beta$ and $\tau$ such that $\alpha -\beta=-\gamma$ and $\alpha
-\tau=-\gamma$ $\Rightarrow$ $\beta=\tau$, thus at a row $\gamma$
if all the entries are not null then there is a single non-zero
entry which is $N_{\alpha,-\beta}$,

\item we should state that the first and the third items are
consistent that is to say if $\alpha -\beta\neq-\gamma$ then this
is valid from either the column or the row point of view,

\item for the diagonal elements $\gamma=\beta$ thus the condition
$\alpha -\beta=-\gamma$ implies that $\alpha=\beta-\gamma=0\notin
\Delta$. However since $\alpha\in \Delta$ the condition $\alpha
-\beta=-\gamma$ can not be held for the diagonal elements
therefore the diagonal elements must all be zero,

\item if $\alpha -\beta\notin\Delta,$ or $\alpha -\beta\in\Delta$
but $ \alpha -\beta\neq-\gamma$ for any $\gamma\in\Delta_{nc}^{+}$
for $n$ times as $\beta$ runs over $\Delta_{nc}^{+}$ then there
are $n$ zero columns. Also since there is a unique non-zero entry
at columns and rows there must be a total number of
dim$\Delta_{nc}^{+}-n$ non-zero entries in dim$\Delta_{nc}^{+}-n$
distinct columns and rows which denotes that there must also be
$n$ zero rows,

\item on the other hand if $\alpha+\gamma\notin\Delta$ for $n$
rows as $\gamma$ runs over $\Delta_{nc}^{+}$ then there are $n$
zero rows. Upon the reasoning given in the previous item there are
also $n$ zero columns. Due to the consistency of the row and the
column point of views these last two items are also consistent.

\end{itemize}

Although as a result of a standard dualization method the
structure constants of the algebra \eqref{eq2} are derived in
\cite{nej1} and \cite{nej2} it is not proven in either of these
works that the algebra defined in \eqref{eq2} which can be called
the dualization deformation of the coset algebra\footnote{Which is
the solvable Lie subalgebra of the global symmetry group of the
sigma model.} of the sigma model at hand forms a Lie algebra.
Therefore in this section we will prove that the algebra defined
in \eqref{eq2} indeed is a Lie algebra. In general an
$m$-dimensional Lie algebra is generated by $m$ generators $X_{a}$
such that
\begin{equation}\label{eq12}
 [X_{a},X_{b}]=C_{\:\:ab}^{c}X_{c},
\end{equation}
with $C_{\:\:ab}^{c}=-C_{\:\:ba}^{c}$ and
\begin{equation}\label{eq14}
 [X_{a},X_{a}]=0.
\end{equation}
The generators must also satisfy the Jacobi identities
\begin{equation}\label{eq15}
 [X_{a},[X_{b},X_{c}]]+[X_{b},[X_{c},X_{a}]]+[X_{c},[X_{a},X_{b}]]=0.
\end{equation}
Thus our task is to show that the generators \eqref{eq1} whose
structure constants are defined in \eqref{eq2} satisfy the Jacobi
identities \eqref{eq15}. At first glance if we choose
$X_{a}=T_{m}$, $X_{b}=T_{n}$, $X_{c}=T_{l}$ \eqref{eq15} is
readily satisfied since the basis $\{T_{m}\}$ generates a Lie
algebra which is the solvable Lie-subalgebra of the Lie algebra of
the global symmetry group of the sigma model Lagrangian
\cite{nej1,nej2}. Next if we choose $X_{a}=\widetilde{T }_{m}$,
$X_{b}=\widetilde{T}_{n}$, $X_{c}=\widetilde{T}_{l}$ from
\eqref{eq6} we get
\begin{equation}\label{eq16}
 [\widetilde{T}_{m},0]+[\widetilde{T}_{n},0]+[\widetilde{T}_{l},0]=0,
\end{equation}
which is also satisfied. If we let $X_{a}=T_{l}$,
$X_{b}=\widetilde{T}_{m}$, $X_{c}=\widetilde{T}_{n}$ then again
from \eqref{eq6} we have
\begin{equation}\label{eq17}
 [T_{l},0]-U_{\:\:ln}^{t}[\widetilde{T}_{m},\widetilde{T}_{t}]+U_{\:\:lm}^{t}[\widetilde{T}_{n},\widetilde{T}_{t}]=0,
\end{equation}
which is instantly satisfied due to the commutation of the dual
generators. In writing \eqref{eq17} we have defined
\begin{equation}\label{eq18}
 [T_{l},\widetilde{T}_{m}]=U_{\:\:lm}^{t}\widetilde{T}_{t},
\end{equation}
where the structure constants $U_{\:\:lm}^{t}$ can be read from
\eqref{eq6}. Now let us consider $X_{a}=T_{l}$, $X_{b}=T_{n}$,
$X_{c}=\widetilde{T}_{m}$. In this case from \eqref{eq15} after
some algebra we find
\begin{equation}\label{eq19}
 Z_{\:\:ln}^{t}U_{\:\:tm}^{s}=(U_{l}U_{n}-U_{n}U_{l})_{\:\:m}^{s},
\end{equation}
where we have defined
\begin{equation}\label{eq20}
[T_{l},T_{n}]=Z_{\:\:ln}^{t}T_{t},
\end{equation}
and we have introduced the structure constant matrices
$(U_{l})_{\:\:m}^{s}=U_{\:\:lm}^{s}$. Now we will prove that for
the three distinct cases;
\begin{enumerate}
\item $T_{l}=H_{j},$$\quad$ $T_{n}=H_{k}$,

\item $T_{l}=H_{j}$,$\quad$ $T_{n}=E_{\gamma}$,

\item $T_{l}=E_{\gamma}$,$\quad$ $T_{n}=E_{\lambda}$,

\end{enumerate}
\eqref{eq19} is satisfied. For the first case $[H_{j},H_{k}]=0$
thus the LHS of \eqref{eq19} vanishes. From \eqref{eq18} and
\eqref{eq6} the RHS of \eqref{eq19} becomes
\begin{equation}\label{eq21}
 (\widetilde{g}_{j}\widetilde{g}_{k}-\widetilde{g}_{k}\widetilde{g}_{j})_{\:\:m}^{s}.
\end{equation}
However from \eqref{eq8} we have
\begin{equation}\label{eq22}
\widetilde{g}_{j}\widetilde{g}_{k}=\overset{\scriptstyle
i\scriptstyle
=\scriptstyle1,\scriptstyle2,\scriptstyle\cdots\cdots\cdot\scriptstyle\scriptstyle
r\quad\quad\quad\quad\:\:\scriptstyle\alpha,\:\:\:\:\:\:\scriptstyle\beta,\scriptstyle\cdots
\scriptstyle\cdots\scriptstyle\cdot\scriptstyle\cdot\scriptstyle\cdot\cdots\quad}
{\left(\begin{array}{c|c} \begin{pmatrix} &&&&\\
&&\text{\Large{0}}&&\\&&&&
\end{pmatrix}&
\begin{pmatrix} &&&&&&
\\&&&\text{\Large{0}} &&&\\&&&&&&\end{pmatrix}    \\
\hline
\begin{pmatrix}&&&&\\&&\text{\Large{0}}&&\\&&&&\end{pmatrix}&
\begin{pmatrix}\alpha_{j}\alpha_{k}&&\text{\Large{0}}\\&\beta_{j}\beta_{k}&\\\text{\Large{0}}&&\ddots\end{pmatrix}
\end{array}\right)}\begin{array}{c}\overset{\overset{\underset{\scriptstyle\shortparallel}{\scriptstyle i}}{
\scriptstyle 1}}{\underset{\scriptstyle
r}{\overset{\scriptstyle\cdot}{\overset{\scriptstyle\cdot}{\scriptstyle\cdot
}}}}\\\underset{\scriptstyle\alpha}{\mathstrut}\\\scriptstyle
\beta\\\scriptstyle\vdots\\\\\end{array},
\end{equation}
and similarly for $\widetilde{g}_{k}\widetilde{g}_{j}$. Thus
inserting these in \eqref{eq21} we see that the RHS of
\eqref{eq19} also vanishes. Now for the second case \eqref{eq19}
becomes
\begin{equation}\label{eq23}
 Z_{\:\:j\gamma}^{t}U_{\:\:tm}^{s}=(U_{j}U_{\gamma}-U_{\gamma}U_{j})_{\:\:m}^{s}.
\end{equation}
After using the identifications
\begin{subequations}\label{eq24}
\begin{gather}
Z_{\:\:j\gamma}^{i}=0\quad,\quad Z_{\:\:j\gamma}^{\beta}=0\quad\text{if}\quad \beta\neq\gamma,\notag\\
\notag\\
Z_{\:\:j\gamma}^{\gamma}=\gamma_{j}\quad,\quad
U_{j}=\widetilde{g}_{j}\quad,\quad
U_{\gamma}=\widetilde{f}_{\gamma},\tag{\ref{eq24}}
\end{gather}
\end{subequations}
in \eqref{eq23} the matrix equality to be proven becomes
\begin{equation}\label{eq25}
 \gamma_{j}\widetilde{f}_{\gamma}\overset{?}{=}\widetilde{g}_{j}\widetilde{f}_{\gamma}-\widetilde{f}_{\gamma}\widetilde{g}_{j}.
\end{equation}
The LHS is
\begin{equation}\label{eq26}
\gamma_{j}\widetilde{f}_{\gamma}=\overset{\scriptstyle
i\scriptstyle
=\scriptstyle1,\scriptstyle2,\scriptstyle\cdots\scriptstyle\cdots\scriptstyle\cdot\scriptstyle
r\quad\quad\quad\scriptstyle\alpha,\scriptstyle\beta,\scriptstyle\cdots\cdots\cdots
\scriptstyle\gamma,\scriptstyle\cdots\scriptstyle\cdot\scriptstyle\cdot\scriptstyle\cdot\cdots\:\:}
{\left(\begin{array}{c|c} \begin{pmatrix} &&&&\\
&&\text{\Large{0}}&&\\&&&&
\end{pmatrix}&
\begin{pmatrix} & \:\:\frac{1}{4}\gamma_{1}\gamma_{j}
\\\text{\Large{0}} &\:\:\vdots&\text{\Large{0}}\\&\:\:\frac{1}{4}\gamma_{r}\gamma_{j} \end{pmatrix}    \\
\hline
\begin{pmatrix}&&&&\\&&\text{\Large{0}}&&\\&&&&\end{pmatrix}&
\begin{pmatrix}&&&&\\&&\gamma_{j}\text{\Large{$R$}}_{\gamma}&&\\&&&&\end{pmatrix}
\end{array}\right)}\begin{array}{c}\overset{\overset{\underset{\scriptstyle\shortparallel}{\scriptstyle
i}}{ \scriptstyle 1}}{\underset{\scriptstyle
r}{\overset{\scriptstyle\cdot}{\overset{\scriptstyle\cdot}{\overset{\scriptstyle\cdot}{\scriptstyle\cdot}
}}}}\\\scriptstyle\alpha\\\scriptstyle
\beta\\\scriptstyle\vdots\\\\\end{array},
\end{equation}
where
\begin{equation}\label{eq27}
\gamma_{j}\text{\large{$R$}}_{\gamma}= \left(\begin{matrix}
\overset{\scriptstyle\alpha}{\mathstrut
0}&\overset{\:\:\scriptstyle\beta}{\mathstrut}&
\overset{\:\:\scriptstyle\cdots}{\mathstrut}&\overset{\scriptstyle\kappa}{\mathstrut
0}&\overset{\scriptstyle\cdots}{\mathstrut}
&\overset{\scriptstyle\cdots}{\mathstrut}&\overset{\scriptstyle\cdots}{\mathstrut}
\\
&\:\:0&&\vdots&&\\&&\:\:\ddots&\vdots&&\\\\&&&0&&\\0&\cdots&0&(\tau,\kappa:=\gamma)\gamma_{j}&0&\cdots&0\\&&&0&\ddots&\\&&&\vdots&&0&
\\&&&0&&&0\end{matrix}\right)
\begin{array}{c}\overset{\scriptstyle\alpha}{\mathstrut}\\\overset{\scriptstyle\beta}{\mathstrut}\\\scriptstyle\vdots\\\scriptstyle\vdots
\\\overset{\scriptstyle\tau}{\mathstrut}\\
\scriptstyle\vdots\\\scriptstyle\vdots\end{array}.
\end{equation}
To calculate the RHS from \eqref{eq7} and \eqref{eq8} we first
find that
\begin{equation}\label{eq28}
\widetilde{g}_{j}\widetilde{f}_{\gamma}=-\overset{\scriptstyle
i\scriptstyle
=\scriptstyle1,\scriptstyle2,\scriptstyle\cdots\scriptstyle\cdots\scriptstyle\cdot\scriptstyle
r\quad\quad\quad\scriptstyle\alpha,\:\:\:\scriptstyle\beta,\scriptstyle\cdots
\scriptstyle\cdots\scriptstyle\cdot\scriptstyle\cdot\scriptstyle\cdot}
{\left(\begin{array}{c|c} \begin{pmatrix} &&&&\\
&&\text{\Large{0}}&&\\&&&&
\end{pmatrix}&
\begin{pmatrix} &&&&
\\&&\text{\Large{0}} &&\\&&&&\end{pmatrix}    \\
\hline
\begin{pmatrix}&&&&\\&&\text{\Large{0}}&&\\&&&&\end{pmatrix}&
\begin{pmatrix}&&\\&\text{\Large{$R$}}_{\gamma j}&\\&&\end{pmatrix}
\end{array}\right)}\begin{array}{c}\overset{\overset{\underset{\scriptstyle\shortparallel}{\scriptstyle i}}{
\scriptstyle 1}}{\underset{\scriptstyle
r}{\overset{\scriptstyle\cdot}{\overset{\scriptstyle\cdot}{\scriptstyle\cdot
}}}}\\\underset{\scriptstyle\alpha}{\mathstrut}\\\scriptstyle
\beta\\\scriptstyle\vdots\\\\\end{array},
\end{equation}
and
\begin{equation}\label{eq29}
\widetilde{f}_{\gamma}\widetilde{g}_{j}=\overset{\scriptstyle
i\scriptstyle
=\scriptstyle1,\scriptstyle2,\scriptstyle\cdots\scriptstyle\cdots\scriptstyle\cdot\scriptstyle
r\quad\quad\quad\scriptstyle\alpha,\scriptstyle\beta,\scriptstyle\cdots\cdots\cdots
\scriptstyle\gamma,\scriptstyle\cdots\scriptstyle\cdot\scriptstyle\cdot\scriptstyle\cdot\cdots\:\:}
{\left(\begin{array}{c|c} \begin{pmatrix} &&&&\\
&&\text{\Large{0}}&&\\&&&&
\end{pmatrix}&
\begin{pmatrix} & \:\:-\frac{1}{4}\gamma_{1}\gamma_{j}
\\\text{\Large{0}} &\:\:\vdots&\text{\Large{0}}\\&\:\:-\frac{1}{4}\gamma_{r}\gamma_{j} \end{pmatrix}    \\
\hline
\begin{pmatrix}&&&&\\&&\text{\Large{0}}&&\\&&&&\end{pmatrix}&
\begin{pmatrix}&&&&\\&&\text{\Large{$R$}}^{\prime}_{\gamma j}&&\\&&&&\end{pmatrix}
\end{array}\right)}\begin{array}{c}\overset{\overset{\underset{\scriptstyle\shortparallel}{\scriptstyle
i}}{ \scriptstyle 1}}{\underset{\scriptstyle
r}{\overset{\scriptstyle\cdot}{\overset{\scriptstyle\cdot}{\overset{\scriptstyle\cdot}{\scriptstyle\cdot}
}}}}\\\scriptstyle\alpha\\\scriptstyle
\beta\\\scriptstyle\vdots\\\\\end{array},
\end{equation}
where we have introduced the matrices
\begin{equation}\label{eq30}
\text{\large{$R$}}_{\gamma j}= \left(\begin{matrix}
\overset{\scriptstyle\alpha}{\mathstrut
0}&\overset{\:\:\scriptstyle\beta}{\mathstrut}&
\overset{\:\:\scriptstyle\cdots}{\mathstrut}&\overset{\scriptstyle\kappa}{\mathstrut
0}&\overset{\scriptstyle\cdots}{\mathstrut}
&\overset{\scriptstyle\cdots}{\mathstrut}&\overset{\scriptstyle\cdots}{\mathstrut}
\\
&\:\:0&&\vdots&&\\&&\:\:\ddots&\vdots&&\\\\&&&0&&\\0&\cdots&0&-(\tau,\kappa:=\gamma)\tau_{j}&0&\cdots&0\\&&&0&\ddots&\\&&&\vdots&&0&
\\&&&0&&&0\end{matrix}\right)
\begin{array}{c}\overset{\scriptstyle\alpha}{\mathstrut}\\\overset{\scriptstyle\beta}{\mathstrut}\\\scriptstyle\vdots\\\scriptstyle\vdots
\\\overset{\scriptstyle\tau}{\mathstrut}\\
\scriptstyle\vdots\\\scriptstyle\vdots\end{array},
\end{equation}
also
\begin{equation}\label{eq31}
\text{\large{$R$}}^{\prime}_{\gamma j}= \left(\begin{matrix}
\overset{\scriptstyle\alpha}{\mathstrut
0}&\overset{\:\:\scriptstyle\beta}{\mathstrut}&
\overset{\:\:\scriptstyle\cdots}{\mathstrut}&\overset{\scriptstyle\kappa}{\mathstrut
0}&\overset{\scriptstyle\cdots}{\mathstrut}
&\overset{\scriptstyle\cdots}{\mathstrut}&\overset{\scriptstyle\cdots}{\mathstrut}
\\
&\:\:0&&\vdots&&\\&&\:\:\ddots&\vdots&&\\\\&&&0&&\\0&\cdots&0&-(\tau,\kappa:=\gamma)\kappa_{j}&0&\cdots&0\\&&&0&\ddots&\\&&&\vdots&&0&
\\&&&0&&&0\end{matrix}\right)
\begin{array}{c}\overset{\scriptstyle\alpha}{\mathstrut}\\\overset{\scriptstyle\beta}{\mathstrut}\\\scriptstyle\vdots\\\scriptstyle\vdots
\\\overset{\scriptstyle\tau}{\mathstrut}\\
\scriptstyle\vdots\\\scriptstyle\vdots\end{array},
\end{equation}
whose non-zero entries coincide in column and row with the
non-zero entries of \eqref{eq27}. Therefore the RHS of
\eqref{eq25} becomes
\begin{equation}\label{eq32}
\widetilde{g}_{j}\widetilde{f}_{\gamma}-\widetilde{f}_{\gamma}\widetilde{g}_{j}=\overset{\scriptstyle
i\scriptstyle
=\scriptstyle1,\scriptstyle2,\scriptstyle\cdots\scriptstyle\cdots\scriptstyle\cdot\scriptstyle
r\quad\quad\quad\quad\scriptstyle\alpha,\scriptstyle\beta,\scriptstyle\cdots\cdots\cdots
\scriptstyle\gamma,\scriptstyle\cdots\scriptstyle\cdot\scriptstyle\cdot\scriptstyle\cdot\cdots\cdots\quad}
{\left(\begin{array}{c|c} \begin{pmatrix} &&&&\\
&&\text{\Large{0}}&&\\&&&&
\end{pmatrix}&
\begin{pmatrix} && \:\:\frac{1}{4}\gamma_{1}\gamma_{j}&
\\&\text{\Large{0}} &\:\:\vdots&\text{\Large{0}}&\\&&\:\:\frac{1}{4}\gamma_{r}\gamma_{j}& \end{pmatrix}    \\
\hline
\begin{pmatrix}&&&&\\&&\text{\Large{0}}&&\\&&&&\end{pmatrix}&
\begin{pmatrix}&&&&\\&&\text{\Large{$R$}}_{\gamma j}-\text{\Large{$R$}}^{\prime}_{\gamma j}&&\\&&&&\end{pmatrix}
\end{array}\right)}\begin{array}{c}\overset{\overset{\underset{\scriptstyle\shortparallel}{\scriptstyle
i}}{ \scriptstyle 1}}{\underset{\scriptstyle
r}{\overset{\scriptstyle\cdot}{\overset{\scriptstyle\cdot}{\overset{\scriptstyle\cdot}{\scriptstyle\cdot}
}}}}\\\scriptstyle\alpha\\\scriptstyle
\beta\\\scriptstyle\vdots\\\\\end{array}.
\end{equation}
We see that from \eqref{eq26} and \eqref{eq32} the three
block-matrices except the one at the lower-rightmost corner are
obviously equal to each other. On the other hand since the
non-zero entries of the matrices \eqref{eq27}, \eqref{eq30} and
\eqref{eq31} coincide we must show that these entries in the
lower-rightmost block-matrices on the LHS and the RHS of
\eqref{eq25} are equal. The non-zero entries of $R_{\gamma
j}-R_{\gamma j}^{\prime}$ at the row $\tau$ and the column
$\kappa$ are
\begin{subequations}\label{eq33}
\begin{align}
-(\tau,\kappa:=\gamma)\tau_{j}+(\tau,\kappa:=\gamma)\kappa_{j}&=-N_{\gamma,-\kappa}\tau_{j}
+N_{\gamma,-\kappa}\kappa_{j}\notag\\
\notag\\
&=N_{\gamma,-\kappa}(\kappa_{j}-\tau_{j}).\tag{\ref{eq33}}
\end{align}
\end{subequations}
However from \eqref{eq10} a non-zero entry exists if and only if
$\gamma-\kappa=-\tau$. This root condition is also valid for the
root vector components and we have
$\kappa_{j}-\tau_{j}=\gamma_{j}$ thus the non-zero entries of
$R_{\gamma j}-R_{\gamma j}^{\prime}$ at the row $\tau$ and the
column $\kappa$ become
\begin{equation}\label{eq34}
N_{\gamma,-\kappa}\gamma_{j}=(\tau,\kappa:=\gamma)\gamma_{j},
\end{equation}
which are equal to the non-zero entries in \eqref{eq27} which is
the LHS lower-rightmost matrix. Finally for the third case of
$T_{l}=E_{\gamma}$, and $T_{n}=E_{\lambda}$ \eqref{eq19} yields
the matrix equality to be proven
\begin{equation}\label{eq35}
 N_{\gamma\lambda}\widetilde{f}_{\gamma+\lambda}\overset{?}{=}\widetilde{f}_{\gamma}\widetilde{f}_{\lambda}
 -\widetilde{f}_{\lambda}\widetilde{f}_{\gamma}.
\end{equation}
In order to obtain this we have used that
\begin{subequations}\label{eq36}
\begin{gather}
Z_{\:\:\gamma\lambda}^{i}=0\quad,\quad Z_{\:\:\gamma\lambda}^{\beta}=0\quad\text{if}\quad \gamma+\lambda\neq\beta,\notag\\
\notag\\
Z_{\:\:\gamma\lambda}^{\gamma+\lambda}=N_{\gamma\lambda}\quad,\quad
U_{\gamma}=\widetilde{f}_{\gamma}.\tag{\ref{eq36}}
\end{gather}
\end{subequations}
We should state that if $\gamma+\lambda$ is not a root then the
LHS is zero. If it is a root then as we have discussed before it
must be in $\Delta_{nc}^{+}$ and in this case the LHS of
\eqref{eq35} becomes
\begin{equation}\label{eq37}
N_{\gamma\lambda}\widetilde{f}_{\gamma+\lambda}=\overset{\scriptstyle
i\scriptstyle
=\scriptstyle1,\scriptstyle2,\scriptstyle\cdots\scriptstyle\cdots\scriptstyle\cdot\scriptstyle
r\quad\quad\quad\scriptstyle\alpha,\scriptstyle\beta,\scriptstyle\cdots\cdots\cdots\cdots
\scriptstyle(\gamma+\lambda),\scriptstyle\cdots\scriptstyle\cdot\scriptstyle\cdot\scriptstyle\cdot\cdots\cdots\cdots\:}
{\left(\begin{array}{c|c} \begin{pmatrix} &&&&\\
&&\text{\Large{0}}&&\\&&&&
\end{pmatrix}&
\begin{pmatrix} & \:\:\frac{1}{4}N_{\gamma\lambda}(\gamma_{1}+\lambda_{1})
\\\text{\Large{0}} &\:\:\vdots&\text{\Large{0}}\\&\:\:\frac{1}{4}N_{\gamma\lambda}(\gamma_{r}+\lambda_{r}) \end{pmatrix}    \\
\hline
\begin{pmatrix}&&&&\\&&\text{\Large{0}}&&\\&&&&\end{pmatrix}&
\begin{pmatrix}&&&&&&\\&&&N_{\gamma\lambda}\text{\Large{$R$}}_{\gamma+\lambda}&&&\\&&&&&&\end{pmatrix}
\end{array}\right)}\begin{array}{c}\overset{\overset{\underset{\scriptstyle\shortparallel}{\scriptstyle
i}}{ \scriptstyle 1}}{\underset{\scriptstyle
r}{\overset{\scriptstyle\cdot}{\overset{\scriptstyle\cdot}{\overset{\scriptstyle\cdot}{\scriptstyle\cdot}
}}}}\\\scriptstyle\alpha\\\scriptstyle
\beta\\\scriptstyle\vdots\\\\\end{array},
\end{equation}
where
\begin{equation}\label{eq38}
\text{\large{$R$}}_{\gamma+\lambda}= \left(\begin{matrix}
\overset{\scriptstyle\alpha}{\mathstrut
0}&\overset{\:\:\scriptstyle\beta}{\mathstrut}&
\overset{\:\:\scriptstyle\cdots}{\mathstrut}&\overset{\scriptstyle\kappa}{\mathstrut
0}&\overset{\scriptstyle\cdots}{\mathstrut}
&\overset{\scriptstyle\cdots}{\mathstrut}&\overset{\scriptstyle\cdots}{\mathstrut}
\\
&\:\:0&&\vdots&&\\&&\:\:\ddots&\vdots&&\\\\&&&0&&\\0&\cdots&0&(\tau,\kappa:=\gamma+\lambda)&0&\cdots&0\\&&&0&\ddots&\\&&&\vdots&&0&
\\&&&0&&&0\end{matrix}\right)
\begin{array}{c}\overset{\scriptstyle\alpha}{\mathstrut}\\\overset{\scriptstyle\beta}{\mathstrut}\\\scriptstyle\vdots\\\scriptstyle\vdots
\\\overset{\scriptstyle\tau}{\mathstrut}\\
\scriptstyle\vdots\\\scriptstyle\vdots\end{array}.
\end{equation}
Now after a non-straightforward matrix multiplication by using
\eqref{eq7} we find that
\begin{equation}\label{eq39}
\widetilde{f}_{\gamma}\widetilde{f}_{\lambda}=\overset{\scriptstyle
i\scriptstyle
=\scriptstyle1,\scriptstyle2,\scriptstyle\cdots\scriptstyle\cdots\scriptstyle\cdot\scriptstyle
r\quad\quad\quad\scriptstyle\alpha,\scriptstyle\beta,\scriptstyle\cdots\scriptstyle\cdots\scriptstyle\cdots
\scriptstyle\cdot\scriptstyle(\gamma+\lambda),\scriptstyle\cdots\scriptstyle\cdots\scriptstyle\cdot\scriptstyle\cdots\scriptstyle\cdot
\:\:}
{\left(\begin{array}{c|c} \begin{pmatrix} &&&&\\
&&\text{\Large{0}}&&\\&&&&
\end{pmatrix}&
\begin{pmatrix} & \:\:\frac{\gamma_{1}}{4}(\gamma,\theta:=\lambda)
\\\text{\Large{0}} &\:\:\vdots&\text{\Large{0}}\\&\:\:\frac{\gamma_{r}}{4}(\gamma,\theta:=\lambda) \end{pmatrix}    \\
\hline
\begin{pmatrix}&&&&\\&&\text{\Large{0}}&&\\&&&&\end{pmatrix}&
\begin{pmatrix}&&&&\\&&\text{\Large{$R$}}_{\gamma\lambda}&&\\&&&&\end{pmatrix}
\end{array}\right)}\begin{array}{c}\overset{\overset{\underset{\scriptstyle\shortparallel}{\scriptstyle
i}}{ \scriptstyle 1}}{\underset{\scriptstyle
r}{\overset{\scriptstyle\cdot}{\overset{\scriptstyle\cdot}{\overset{\scriptstyle\cdot}{\scriptstyle\cdot}
}}}}\\\scriptstyle\alpha\\\scriptstyle
\beta\\\scriptstyle\vdots\\\\\end{array},
\end{equation}
where $\theta=\gamma+\lambda$ and if $\gamma+\lambda$ is not a
root then the upper-rightmost block matrix in \eqref{eq39} is zero
since in this case there would be no column $\theta$ in
$R_{\lambda}$ which would satisfy $\theta=\gamma+\lambda$ so that
the $\gamma$ row of $R_{\lambda}$ would be composed of all zero
elements. In \eqref{eq39} we have defined
\begin{equation}\label{eq40}
\text{\large{$R$}}_{\gamma\lambda}=\text{\large{$R$}}_{\gamma}\text{\large{$R$}}_{\lambda},
\end{equation}
which through \eqref{eq9} can explicitly be calculated as
\begin{equation}\label{eq41}
\text{\large{$R$}}_{\gamma\lambda}= \left(\begin{matrix}
\overset{\scriptstyle\alpha}{\mathstrut
0}&\overset{\:\:\quad\quad\scriptstyle\beta}{\mathstrut}&
\overset{\:\:\quad\quad\quad\scriptstyle\cdots\scriptstyle\cdots\scriptstyle\cdots
\scriptstyle\cdots\scriptstyle\cdots}{\mathstrut}&\overset{\scriptstyle\kappa}
{\mathstrut 0}&\overset{\scriptstyle\cdots}{\mathstrut}
&\overset{\scriptstyle\cdots}{\mathstrut}&\overset{\scriptstyle\cdots}{\mathstrut}
\\
&\quad\quad
0&&\vdots&&\\&&&\vdots&&\\&&\quad\quad\ddots&\vdots&&\\&&&\vdots&&\\\\&&&0&&\\0&\quad\cdots&0&(\upsilon,\kappa:=\lambda)
(\tau,\upsilon:=\gamma)&0&\cdots&0\\&&&0&&\\&&&\vdots&&\ddots&
\\&&&0&&&0\end{matrix}\right)
\begin{array}{c}\scriptstyle\alpha\\\\
\overset{\scriptstyle\beta}{\mathstrut}\\\scriptstyle\vdots\\\scriptstyle\vdots\\\scriptstyle\vdots\\
\\\overset{\scriptstyle\tau}{\mathstrut}\\
\scriptstyle\vdots\\\scriptstyle\vdots\end{array}.
\end{equation}
In calculating $R_{\gamma\lambda}$ we have efficiently made use of
the properties of $R_{\alpha}$ which we have itemized before.
Apart from the values of its non-zero entries and where they are
the matrix \eqref{eq41} also obeys all the characteristics of
$R_{\alpha}$. That is to say its diagonal elements are zero, it
may have columns or rows whose elements are all zero but their
multiplicity must be equal, only one non-zero element can exist in
a row also only one non-zero element can exist in a column. The
orientation of an entry at a column $\kappa$ in \eqref{eq41} can
be found as follows; one first solves the non-zero entry condition
$\lambda-\kappa=-\upsilon$ of $R_{\lambda}$ for $\upsilon$ then
one solves the row $\tau$ from the non-zero entry condition
$\gamma-\upsilon=-\tau$ of $R_{\gamma}$. This is because in the
matrix multiplication in \eqref{eq40} we multiply the unique
non-zero element in the column $\kappa$ which is at the row
$\upsilon$ of $R_{\lambda}$ with the unique non-zero element in
the column $\upsilon$ which is at the row $\tau$ of $R_{\gamma}$
and write it in the column $\kappa$ and the row $\tau$ of
$R_{\gamma\lambda}$. Of course if the condition
$\lambda-\kappa=-\upsilon$ is not satisfied for any $\upsilon$
then this generates a pair of a column and a row which are both
null in $R_{\lambda}$ and the $\kappa$ column of
$R_{\gamma\lambda}$ would be null too. Also separately if the
condition $\gamma-\upsilon=-\tau$ is not satisfied for any $\tau$
then this generates a pair of a column and a row which are also
null in $R_{\gamma}$ and the $\kappa$ column of
$R_{\gamma\lambda}$ would again be null. These two cases may
coexist. In addition in either of these cases there would also be
a completely null row in $R_{\gamma\lambda}$. Now we can write the
RHS of \eqref{eq35} as
\begin{equation}\label{eq42}
\widetilde{f}_{\gamma}\widetilde{f}_{\lambda}-\widetilde{f}_{\lambda}\widetilde{f}_{\gamma}=\overset{\scriptstyle
i\scriptstyle
=\scriptstyle1,\scriptstyle2,\scriptstyle\cdots\scriptstyle\cdots\scriptstyle\cdot\scriptstyle
r\quad\quad\quad\scriptstyle\alpha,\scriptstyle\beta,
\scriptstyle\cdots\scriptstyle\cdots\scriptstyle\cdots\scriptstyle\cdots\scriptstyle\cdots
\scriptstyle\cdot\scriptstyle(\gamma+\lambda),\scriptstyle\cdots\scriptstyle\cdots\scriptstyle\cdot
\scriptstyle\cdots\scriptstyle\cdots \:\:\quad}
{\left(\begin{array}{c|c} \begin{pmatrix} &&&&\\
&&\text{\Large{0}}&&\\&&&&
\end{pmatrix}&
\begin{pmatrix} &
\:\:\frac{\gamma_{1}(\gamma,\theta:=\lambda)-\lambda_{1}(\lambda,\theta:=\gamma)}{4}
\\\text{\Large{0}} &\:\:\vdots&\text{\Large{0}}\\&\:\:\frac{\gamma_{r}
(\gamma,\theta:=\lambda)-\lambda_{r}(\lambda,\theta:=\gamma)}{4}\end{pmatrix}    \\
\hline
\begin{pmatrix}&&&&\\&&\text{\Large{0}}&&\\&&&&\end{pmatrix}&
\begin{pmatrix}&&&&\\&&\text{\Large{$R$}}_{\gamma\lambda}-\text{\Large{$R$}}_{\lambda\gamma}&&\\&&&&\end{pmatrix}
\end{array}\right)}\begin{array}{c}\overset{\overset{\underset{\scriptstyle\shortparallel}{\scriptstyle
i}}{ \scriptstyle 1}}{\underset{\scriptstyle
r}{\overset{\scriptstyle\cdot}{\overset{\scriptstyle\cdot}{\overset{\scriptstyle\cdot}{\scriptstyle\cdot}
}}}}\\\scriptstyle\alpha\\\scriptstyle
\beta\\\scriptstyle\vdots\\\\\end{array}.
\end{equation}
Before going further we should show that the non-zero entries of
the matrices $R_{\gamma\lambda}$ and $R_{\lambda\gamma}$ indeed
coincide. This can be seen as follows; if there is a non-zero
entry at a row $\tau$ in $R_{\gamma\lambda}$ this means that
\begin{equation}\label{eq43}
 \lambda-\kappa=-\upsilon,\quad\text{and}\quad\gamma-\upsilon=-\tau,
\end{equation}
if one adds these two root conditions side by side one finds that
the non-zero entry must be at the column
$\kappa=\gamma+\lambda+\tau$ which is certainly an element of
$\Delta_{nc}^{+}$, also if there is a non-zero entry at a row
$\tau$ in $R_{\lambda\gamma}$ then we have
\begin{equation}\label{eq44}
 \gamma-\kappa=-\xi,\quad\text{and}\quad\lambda-\xi=-\tau,
\end{equation}
thus again adding side by side gives us the column of the non-zero
entry which also becomes $\kappa=\gamma+\lambda+\tau$. Therefore
if both matrices have a non-zero entry at a row (which are unique)
their difference also has a non-zero entry at that row which is
also unique. If either $R_{\gamma\lambda}$ or $R_{\lambda\gamma}$
has a zero row then again $R_{\gamma\lambda}-R_{\lambda\gamma}$
has a unique non-zero entry at that row. Also if both of the
matrices $R_{\gamma\lambda}$ and $R_{\lambda\gamma}$ have
coinciding zero-rows then $R_{\gamma\lambda}-R_{\lambda\gamma}$
will have zero elements in that row. These facts show us that the
matrix $R_{\gamma\lambda}-R_{\lambda\gamma}$ has zero diagonal
elements and if it has a non-zero entry at a row then that entry
must be unique. However on the other hand
$R_{\gamma\lambda}-R_{\lambda\gamma}$ may have zero rows too. The
matrix $R_{\gamma\lambda}-R_{\lambda\gamma}$ also has unique
non-zero entries at its columns if they exist. This is due to two
facts; firstly if both matrices have non-zero entries at a row
then  as we have discussed above they coincide and since both
matrices have unique entries in a column their difference will
have a unique entry at the corresponding column, secondly if one
of the matrices has a zero $\tau$ row but the other's $\tau$ row
is not zero, the one which has a non-zero $\tau$ row will have a
unique entry at the column $\kappa=\gamma+\lambda+\tau$, also it
can be seen from the conditions \eqref{eq43} and \eqref{eq44} that
the other one which has the zero $\tau$ row must have zero
elements in the column $\kappa=\gamma+\lambda+\tau$ as if it has a
non-zero element in the column $\kappa=\gamma+\lambda+\tau$ at a
different row say $\phi$ than $\tau$ this would imply
$\kappa=\gamma+\lambda+\phi$ which would contradict with
$\kappa=\gamma+\lambda+\tau$ which is obtained through the
addition of the non-zero-entry existence conditions of the first
matrix whose $\tau$ row is not composed of zero elements. On the
other hand $\tau=\phi$ would contradict with the assumption that
the second matrix in question has zero $\tau$ row. Thus its
$\kappa=\gamma+\lambda+\tau$ column must be a null-column. These
two facts denote that if there is a non-zero entry in a column of
$R_{\gamma\lambda}-R_{\lambda\gamma}$ then it must be unique. For
this reason if $R_{\gamma\lambda}-R_{\lambda\gamma}$ has $n$ zero
rows then it also has $n$ zero columns. We immediately see that
the block matrices $R_{\gamma+\lambda}$ and
$R_{\gamma\lambda}-R_{\lambda\gamma}$ which are on the LHS and the
RHS of \eqref{eq35} respectively obey similar properties like zero
diagonal elements and unique non-zero row or column entries if
they exist. If at a fixed row $\tau$ there is a non-zero entry
(which is unique) in $R_{\gamma+\lambda}$ then from \eqref{eq38}
we deduce that it must be at the column
$\kappa=\gamma+\lambda+\tau$. Also following our discussion above
at a fixed row $\tau$ if $R_{\gamma\lambda}-R_{\lambda\gamma}$ has
a non-zero entry which is also unique then it must also be at the
column $\kappa=\gamma+\lambda+\tau$ too\footnote{Again from our
discussion about the structure of
$R_{\gamma\lambda}-R_{\lambda\gamma}$ we know that this is valid
at a row $\tau$ for either of the cases when both
$R_{\gamma\lambda}$
 and $R_{\lambda\gamma}$ contribute a non-zero entry, and when only one of them
 contributes.}. On the other hand at a fixed row $\tau$ if
 $R_{\gamma+\lambda}$ does not have a non-zero entry this means
 that there exists no column $\kappa$ which would satisfy
 $\gamma+\lambda-\kappa=-\tau$\footnote{This can happen either if
 for none of the columns $\kappa$,
 $\gamma+\lambda-\kappa\in\Delta$ or $\gamma+\lambda-\kappa\neq-\tau$ if there exists a $\kappa$
such that  $\gamma+\lambda-\kappa\in\Delta$.}. In this case there
can not be a non-zero entry at the row $\tau$ on the RHS in
$R_{\gamma\lambda}-R_{\lambda\gamma}$ as if it exists from either
\eqref{eq43} or \eqref{eq44} we must have
$\kappa=\gamma+\lambda+\tau$ which would contradict with the
assumed impossibility of this root condition at the $\tau$ row of
$R_{\gamma+\lambda}$ on the LHS. In summary we conclude that: $i)$
if both $R_{\gamma+\lambda}$ and
$R_{\gamma\lambda}-R_{\lambda\gamma}$ have a non-zero entry at a
row $\tau$ they must coincide, $ii)$ if $R_{\gamma+\lambda}$ does
not have a non-zero entry at a row $\tau$ then
$R_{\gamma\lambda}-R_{\lambda\gamma}$ can not have a non-zero
entry at the same row. Therefore all the entries of
$R_{\gamma+\lambda}$ and $R_{\gamma\lambda}-R_{\lambda\gamma}$
coincide and we may question the equality of the lower rightmost
block matrices on the LHS and the RHS of \eqref{eq35} whose
entries coincide. However we should state that we leave dealing
with the case of a non-zero entry at a row $\tau$ in
$R_{\gamma+\lambda}$ but all zero entries at the row $\tau$ of
$R_{\gamma\lambda}-R_{\lambda\gamma}$ for later. We will prove
that in this case the entry on the LHS must be also zero due to
structure constant conditions of the Cartan-Weyl basis.

Now if we take a look at the non-zero entries of the
upper-rightmost block matrix of \eqref{eq42} which are on the
column $\gamma+\lambda$ we have
\begin{equation}\label{eq45}
\frac{\gamma_{i}(\gamma,\theta:=\lambda)-\lambda_{i}(\lambda,\theta:=\gamma)}{4}
=\frac{\gamma_{i}N_{\lambda,-\theta}-\lambda_{i}N_{\gamma,-\theta}}{4}.
\end{equation}
As we have discussed before if $\gamma+\lambda$ is not a root then
the upper-rightmost block matrix of \eqref{eq42} will be zero
which will be equal to the upper-rightmost block matrix of the LHS
of \eqref{eq35} which will again be zero owing to the vanishing of
$N_{\gamma\lambda}$. However if $\gamma+\lambda$ is a root then
through our previous discussion about the closure of the coset
algebra it must be in $\Delta_{nc}^{+}$ and then there exists a
root $\gamma+\lambda=\theta\in\Delta_{nc}^{+}$ such that
$\lambda-\theta=-\gamma$ and $\gamma-\theta=-\lambda$. Therefore
in this case the non-zero entries given in \eqref{eq45} do exist.
For the root generators of a Cartan-Weyl basis if
$\alpha+\beta+\gamma=0$ then we have
\begin{equation}\label{eq46}
N_{\alpha\beta}=N_{\beta\gamma}=N_{\gamma\alpha}.
\end{equation}
Since for the non-zero entries \eqref{eq45}
$\gamma+\lambda-\theta=0$ we have
$N_{\lambda,-\theta}=N_{\gamma\lambda}$ and
$N_{\gamma,-\theta}=-N_{-\theta,\gamma}=-N_{\gamma\lambda}$. Thus
the non-zero entries in \eqref{eq45} become
\begin{equation}\label{eq47}
\frac{\gamma_{i}N_{\lambda,-\theta}-\lambda_{i}N_{\gamma,-\theta}}{4}=
\frac{N_{\gamma\lambda}(\gamma_{i}+\lambda_{i})}{4},
\end{equation}
which are equal to the non-zero entries in the upper-rightmost
block matrix of \eqref{eq37} which are also on the column
$\gamma+\lambda$.

Now we will come back to the question of the equality of the lower
rightmost block matrices on the LHS and the RHS of \eqref{eq35}.
Firstly let us assume that $\gamma+\lambda$ is not a root as we
have mentioned before in this case directly from \eqref{eq15} the
LHS of \eqref{eq35} is zero. If on the RHS in
$R_{\gamma\lambda}-R_{\lambda\gamma}$ at a row $\tau$ and a column
$\kappa$ both of the root conditions \eqref{eq43} and \eqref{eq44}
hold and an entry exists then it must be
\begin{equation}\label{eq48}
N_{\lambda,-\kappa}N_{\gamma,-\upsilon}-N_{\gamma,-\kappa}N_{\lambda,-\xi}.
\end{equation}
However from the root conditions \eqref{eq43} and \eqref{eq44} of
entry existence, also from the identity \eqref{eq46} we have
\begin{equation}\label{eq49}
N_{\gamma,-\upsilon}=N_{\tau\gamma}\quad,\quad
N_{\lambda,-\xi}=N_{\tau\lambda}.
\end{equation}
Thus the entry becomes
\begin{equation}\label{eq50}
N_{\lambda,-\kappa}N_{\tau\gamma}-N_{\gamma,-\kappa}N_{\tau\lambda}.
\end{equation}
In general if for four roots $\alpha+\beta+\gamma+\delta=0$ and if
none of the pairs sum up to zero then the Cartan-Weyl basis
structure constants obey the identity
\begin{equation}\label{eq51}
N_{\alpha\beta}N_{\gamma\delta}+N_{\beta\gamma}N_{\alpha\delta}+N_{\gamma\alpha}N_{\beta\delta}=0.
\end{equation}
We have shown that if one adds the entry existence conditions
\eqref{eq43} and \eqref{eq44} side by side one gets the root
condition $\gamma+\lambda+\tau-\kappa=0$. Now if we apply
\eqref{eq51} then we have
\begin{equation}\label{eq52}
N_{\gamma\lambda}N_{\tau,-\kappa}+N_{\lambda\tau}N_{\gamma,-\kappa}+N_{\tau\gamma}N_{\lambda,-\kappa}=0.
\end{equation}
However since we assume the case when $\gamma+\lambda$ is not a
root $N_{\gamma\lambda}=0$ and we have
\begin{equation}\label{eq53}
-N_{\tau\lambda}N_{\gamma,-\kappa}+N_{\tau\gamma}N_{\lambda,-\kappa}=0,
\end{equation}
whose LHS is exactly equal to \eqref{eq50}. This proves the
equality of the lower rightmost block matrices on the LHS and the
RHS of \eqref{eq35} when $\gamma+\lambda$ is not a root and when
\eqref{eq43} and \eqref{eq44} both hold. As a second case when
$\gamma+\lambda$ is not a root if at least one of the root
conditions is not satisfied in both \eqref{eq43} and \eqref{eq44}
then of course the lower rightmost block matrix on the RHS of
\eqref{eq35} will be zero as well as the LHS one. On the other
hand if both of the root conditions are satisfied in \eqref{eq43}
but at least one condition is not satisfied in \eqref{eq44} then
again we have $\gamma+\lambda+\tau-\kappa=0$ and the identity
\eqref{eq52} holds. This shows that a case in which one of the
conditions holds but the other one does not hold in \eqref{eq44}
would be contradictory as in this case one can take the difference
of $\gamma+\lambda+\tau-\kappa=0$ with the holding root condition
to show that the second condition in \eqref{eq44} must also hold.
Now $\gamma+\lambda+\tau-\kappa=0$ can be written as
$\gamma-\kappa=-(\lambda+\tau)$. This shows that if
$\gamma-\kappa\in \Delta$ then $-(\lambda+\tau)\in\Delta$ also
$\lambda+\tau\in\Delta$. Since $\lambda,\tau\in\Delta_{nc}^{+}$ we
have $\lambda+\tau\in\Delta_{nc}^{+}$. Thus it can not be true
that if $\gamma-\kappa\in \Delta$ there does not exist any
$\xi\in\Delta_{nc}^{+}$ which would satisfy $\gamma-\kappa=-\xi$
in \eqref{eq44} as in this case $\xi$ is nothing but
$\lambda+\tau$. Therefore when \eqref{eq43} holds the only
possible conditions of the non-existence of \eqref{eq44} are;
$\gamma-\kappa\notin \Delta$ and $\lambda-\xi\notin \Delta$ or
$\lambda-\xi\in \Delta$ but there exists no
$\tau\in\Delta_{nc}^{+}$ such that $\lambda-\xi=-\tau$. Thus in
this case the entry in $R_{\gamma\lambda}-R_{\lambda\gamma}$ at
the row $\tau$ and the column $\kappa$ becomes
\begin{equation}\label{eq54}
N_{\lambda,-\kappa}N_{\gamma,-\upsilon}.
\end{equation}
Since \eqref{eq43} holds from the identities \eqref{eq46} we again
have $N_{\gamma,-\upsilon}=N_{\tau\gamma}$ so that \eqref{eq54}
can be written as
\begin{equation}\label{eq55}
N_{\lambda,-\kappa}N_{\tau\gamma}.
\end{equation}
However since $N_{\gamma,-\kappa}=0$ and we assume that
$\gamma+\lambda$ is not a root giving $N_{\gamma\lambda}=0$ from
\eqref{eq52} we have
\begin{equation}\label{eq56}
N_{\tau\gamma}N_{\lambda,-\kappa}=0,
\end{equation}
which proves the equality of the lower rightmost block matrices on
the LHS and the RHS of \eqref{eq35} when $\gamma+\lambda$ is not a
root and when \eqref{eq43} holds but \eqref{eq44} does not hold.
Now instead if \eqref{eq44} holds but \eqref{eq43} does not hold
then a similar reasoning and analysis denotes that in this case
the only possible conditions of the non-existence of \eqref{eq43}
are; $\lambda-\kappa\notin \Delta$ and $\gamma-\upsilon\notin
\Delta$ or $\gamma-\upsilon\in \Delta$ but there exists no
$\tau\in\Delta_{nc}^{+}$ such that $\gamma-\upsilon=-\tau$. Thus
in this case the entry in $R_{\gamma\lambda}-R_{\lambda\gamma}$ at
the row $\tau$ and the column $\kappa$ is
\begin{equation}\label{eq57}
-N_{\gamma,-\kappa}N_{\lambda,-\xi}.
\end{equation}
The conditions in \eqref{eq44} hold thus again from the identities
\eqref{eq46} we have $N_{\lambda,-\xi}=N_{\tau\lambda}$ so that
\eqref{eq57} becomes
\begin{equation}\label{eq58}
-N_{\gamma,-\kappa}N_{\tau\lambda}.
\end{equation}
In this case since $N_{\lambda,-\kappa}=0$ and again
$\gamma+\lambda$ is not a root giving $N_{\gamma\lambda}=0$ from
\eqref{eq52} we have
\begin{equation}\label{eq59}
-N_{\tau\lambda}N_{\gamma,-\kappa}=0.
\end{equation}
By considering all the possible cases we have completed the proof
of the equality of the lower rightmost block matrices on the LHS
and the RHS of \eqref{eq35} when $\gamma+\lambda$ is not a root.
Our next task will be to perform a similar proof for the case when
$\gamma+\lambda$ is a root. We have already mentioned that if
$R_{\gamma+\lambda}$ does not have a non-zero entry at a row
$\tau$ then $R_{\gamma\lambda}-R_{\lambda\gamma}$ can not have a
non-zero entry at the same row and we have shown that the entries
of $R_{\gamma+\lambda}$ and $R_{\gamma\lambda}-R_{\lambda\gamma}$
coincide. Thus for the following we will assume that there exists
an entry at the row $\tau$ in $R_{\gamma+\lambda}$ which means
that the root condition $\gamma+\lambda-\kappa=-\tau$ holds due to
\eqref{eq38}. Thus in this case again \eqref{eq52} is valid. We
will start with the case in which the root conditions in
\eqref{eq43} and \eqref{eq44} are both satisfied so that
$R_{\gamma\lambda}$ and $R_{\lambda\gamma}$ both have entries at
the row $\tau$ and the column $\kappa$. Then from \eqref{eq35} at
the row $\tau$ and the column $\kappa$ the equality of the
coinciding entries of the lower rightmost block matrices on the
LHS and the RHS to be proven becomes
\begin{equation}\label{eq60}
 N_{\gamma\lambda}N_{\gamma+\lambda,-\kappa}\overset{?}{=}
 N_{\lambda,-\kappa}N_{\gamma,-\upsilon}-N_{\gamma,-\kappa}N_{\lambda,-\xi}.
\end{equation}
Existence of the root conditions \eqref{eq43} and \eqref{eq44}
again allows the usage of the identity \eqref{eq46} and
\eqref{eq60} becomes
\begin{equation}\label{eq61}
 N_{\gamma\lambda}N_{\gamma+\lambda,-\kappa}\overset{?}{=}
 N_{\lambda,-\kappa}N_{\tau\gamma}-N_{\gamma,-\kappa}N_{\tau\lambda}.
\end{equation}
Now since $\gamma+\lambda-\kappa+\tau=0$ from \eqref{eq46} we have
\begin{equation}\label{eq62}
 N_{\gamma+\lambda,-\kappa}=N_{-\kappa,\tau}.
\end{equation}
Therefore \eqref{eq61} can be written as
\begin{equation}\label{eq63}
 0\overset{?}{=}N_{\gamma\lambda}N_{\tau,-\kappa}
 +N_{\lambda,-\kappa}N_{\tau\gamma}+N_{\gamma,-\kappa}N_{\lambda\tau}.
\end{equation}
However this equality holds due to the validity of \eqref{eq52}
which is the desired result. The next step is to show that the
equality in \eqref{eq60} holds when the root conditions
\eqref{eq43} and \eqref{eq44} are partially satisfied or not
satisfied at all\footnote{This is a case which we have postponed
to deal with before.}. If we refer to our previous root condition
analysis which we have done for \eqref{eq43} and \eqref{eq44} when
we have discussed the cases when $\gamma+\lambda$ is not a root we
can conclude that the following three cases are the only ones
which are not contradictory with the condition
$\gamma+\lambda-\kappa+\tau=0$ which comes from the existence of
the lower rightmost block matrix entry on the LHS of \eqref{eq35};
\begin{itemize}
\item both of the root conditions in \eqref{eq43} hold, in
addition $\gamma-\kappa\notin \Delta$, and $\lambda-\xi\notin
\Delta$ or $\lambda-\xi\in \Delta$ but there exists no
$\tau\in\Delta_{nc}^{+}$ such that $\lambda-\xi=-\tau$,

 \item both of the root conditions in \eqref{eq44} hold, in addition
 $\lambda-\kappa\notin \Delta$, and $\gamma-\upsilon\notin
\Delta$ or $\gamma-\upsilon\in \Delta$ but there exists no
$\tau\in\Delta_{nc}^{+}$ such that $\gamma-\upsilon=-\tau$,

\item $\gamma-\kappa\notin \Delta$, and $\lambda-\xi\notin \Delta$
or $\lambda-\xi\in \Delta$ but there exists no
$\tau\in\Delta_{nc}^{+}$ such that $\lambda-\xi=-\tau$, in
addition $\lambda-\kappa\notin \Delta$, and $\gamma-\upsilon\notin
\Delta$ or $\gamma-\upsilon\in \Delta$ but there exists no
$\tau\in\Delta_{nc}^{+}$ such that $\gamma-\upsilon=-\tau$.
\end{itemize}
For the first case we have to question
\begin{equation}\label{eq64}
 N_{\gamma\lambda}N_{\gamma+\lambda,-\kappa}\overset{?}{=}
 N_{\lambda,-\kappa}N_{\gamma,-\upsilon}.
\end{equation}
However this equation is the same with \eqref{eq60} if we use
$N_{\gamma,-\kappa}=0$ in \eqref{eq60} which is the characteristic
feature of the first case. Bearing in mind the identity
$N_{\gamma,-\upsilon}=N_{\tau\gamma}$ (since \eqref{eq43} is
satisfied) for this first case the proof of the equality in
\eqref{eq64} coincides with the one we have performed for the
previous case following \eqref{eq60}. The other symmetrical case
namely the second one leads to
\begin{equation}\label{eq65}
 N_{\gamma\lambda}N_{\gamma+\lambda,-\kappa}\overset{?}{=}
 -N_{\gamma,-\kappa}N_{\lambda,-\xi}.
\end{equation}
Again this equation is the same with \eqref{eq60} if one lets
$N_{\lambda,-\kappa}=0$ in \eqref{eq60} which is the
characteristic feature of the second case. Upon the insertion of
$N_{\lambda,-\xi}=N_{\tau\lambda}$ (since \eqref{eq44} is
satisfied) the proof of \eqref{eq65} again coincides with the one
following \eqref{eq60}. For the last item which is the combination
of the first and the second ones one has to show
\begin{equation}\label{eq66}
 N_{\gamma\lambda}N_{\gamma+\lambda,-\kappa}\overset{?}{=}0.
\end{equation}
One can obtain this equality from \eqref{eq60} by using
$N_{\gamma,-\kappa}=0$ and $N_{\lambda,-\kappa}=0$ which are the
characteristics of the third case. Thus also for this case the
proof of the equality of \eqref{eq66} comes automatically from the
previous one following \eqref{eq60}. Therefore with this last case
we have shown that when a non-zero entry at a row $\tau$ in
$R_{\gamma+\lambda}$ exists but the $\tau$ row of
$R_{\gamma\lambda}-R_{\lambda\gamma}$ consists of zero elements
the entry on the LHS of \eqref{eq35} must be also zero due to
structure constant conditions of the Cartan-Weyl basis. By this we
have completed the proof of the equality in \eqref{eq35} for all
the possible cases which may arise. In conclusion, we can state
that together with our previous results we have proven that the
algebra structure given in \eqref{eq2} which is a deformation of
the solvable Lie subalgebra of the global symmetry group of the
sigma model obeys the Jacobi identities \eqref{eq15} thus it
defines a Lie algebra.
\section{The Adjoint Representation}
In this section we will show that the dualized coset algebra given
in \eqref{eq2} contains an adjoint representation for the
subalgebra $s$ which is generated by the original coset generators
$\{T_{m}\}\equiv\{H_{i},E_{\alpha}\}$. This subalgebra is nothing
but the original coset algebra of the sigma model which is the
solvable subalgebra of the Lie algebra of the global symmetry
group. The adjoint representation we mention exists due to the
general scheme
\begin{equation}\label{eq67}
[\{T_{m}\},\{\widetilde{T}_{n}\}]\subset\{\widetilde{T}_{n}\},
\end{equation}
of the structure of the dualized coset algebra \eqref{eq2}. Now to
display this representation let us consider the linear\footnote{As
we define it.} map
\begin{equation}\label{eq68}
f\: :\:s\longrightarrow gl(S,\Bbb{R}),
\end{equation}
where $S$ is the dimension of $s$. Similar to the general adjoint
representation of a generic Lie algebra we assume $f$ is such that
\begin{equation}\label{eq69}
f(T_{l})=U_{l},
\end{equation}
where $U_{l}$ is the $S\times S$ matrix whose entries are
$U^{t}_{\:\: lm}$ which are the structure constants defined in
\eqref{eq18}. Linearity and \eqref{eq69} defines the action of $f$
on entire $s$. This linear map becomes an algebra homomorphism if
\begin{equation}\label{eq70}
f([M,N])=[f(M),f(N)],
\end{equation}
for all $M,N\in s$. If one inserts $M=M^{l}T_{l}$ and
$N=N^{k}T_{k}$ in \eqref{eq70} one sees that \eqref{eq70} holds if
\begin{equation}\label{eq71}
f([T_{l},T_{n}])=[f(T_{l}),f(T_{n})].
\end{equation}
By using \eqref{eq20}, and \eqref{eq69}, also the fact that $f$ is
assumed to be a linear map the equality \eqref{eq71} which is in
question can be written as
\begin{equation}\label{eq72}
 Z_{\:\:ln}^{t}U_{\:\:tm}^{s}\overset{?}{=}(U_{l}U_{n}-U_{n}U_{l})_{\:\:m}^{s}.
\end{equation}
However this equality is the same with the Jacobi identity
\eqref{eq19} which we have exactly proven to hold when we showed
that the dualized coset algebra \eqref{eq2} is a Lie algebra in
the previous section. Thus we can conclude that $f$ whose action
on the basis $\{T_{m}\}$ is defined via \eqref{eq69} is an algebra
homomorphism and it forms a $S\times S$ matrix representation for
the coset algebra $s$ which is a subalgebra in \eqref{eq2}. We may
state that the dualized coset algebra \eqref{eq2} which is a
deformation of its subalgebra $s$ and whose Lie algebra structure
is proven in the previous section generates a natural adjoint
representation for the original coset algebra $s$.
\section{The First-order Sigma Model Field Equations}
We will now show that the representation presented in the last
section enables one to derive the first-order field equations of
the symmetric space sigma model. The first-order field equations
of the sigma models with symmetric space coset target manifolds
firstly appeared in \cite{nej1,nej2} as a result of the dualized
coset construction of these theories. However they were formally
generated in those works as consistency conditions within the
dualization of the theory. Here we will algebraically prove that
they correspond to the equations which would be obtained by a
local integration of second-order field equations. In other words
we will show that if one chooses the representation mentioned in
the previous section then one can obtain the second-order field
equations of the symmetric space sigma model by taking the
exterior derivative of the first-order ones derived in
\cite{nej1,nej2}. Thus our starting point is adopting from
\cite{nej1,nej2} the set of equations
\begin{equation}\label{eq73}
\ast \overset{\rightharpoonup }{\mathbf{\Psi }}=(-1)^{D}e^{\mathbf{\Gamma }%
}e^{\mathbf{\Lambda }}\overset{\rightharpoonup }{\mathbf{A}},
\end{equation}
where $D$ is the dimension of the base manifold and the
S-dimensional column vectors $\overset{\rightharpoonup
}{\mathbf{\Psi }}$ and $\overset{\rightharpoonup }{\mathbf{A}}$
have the components
\begin{subequations}\label{eq74}
\begin{gather}
\mathbf{\Psi}^{i}=\frac{1}{2}d\phi^{i},\quad \text{for}\quad
i=1,...,r,\notag\\
\notag\\
\mathbf{\Psi}^{\alpha+r}=e^{\frac{1}{2}\alpha_{i}\phi^{i}}\mathbf{\Omega
}_{\:\:\:\gamma}^{\alpha}d\chi^{\gamma},\quad\text{for}\quad\alpha=1,...,S-r,\notag\\
\notag\\
\mathbf{A}^{i}=\frac{1}{2}d\widetilde{\phi}^{i},\quad\text{for}\quad
i=1,...,r,\quad\text{and}\quad
\mathbf{A}^{\alpha+r}=d\widetilde{\chi}^{\alpha},\quad\text{for}\quad\alpha=1,...,S-r.
 \tag{\ref{eq74}}
\end{gather}
\end{subequations}
Here $\phi^{i}$ and $\chi^{\gamma}$ are the scalar fields to be
solved which parametrize the coset space target manifold of the
sigma model and we should state that $\alpha$ stands both for the
non-compact positive roots and their corresponding enumeration.
Also $\widetilde{\phi}^{i}$ and $\widetilde{\chi}^{\alpha}$ are
arbitrary $(D-2)$-forms which emerge from the dualization of the
coset map within the dualized coset realization of the theory. In
\eqref{eq73} $\mathbf{\Gamma }(\phi^{i})$ and $\mathbf{\Lambda
}(\chi^{\beta})$ are S$\times$S matrix functionals with components
\begin{equation}\label{eq75}
\mathbf{\Gamma }%
_{n}^{k}=\frac{1}{2}\phi ^{i}\,\widetilde{g}_{in}^{k}\quad,\quad
\mathbf{\Lambda }_{n}^{k}=\chi ^{\alpha}\widetilde{f}_{\alpha
n}^{k},
\end{equation}
where the matrices $\widetilde{f}_{\alpha}$ and
$\widetilde{g}_{i}$ are defined in \eqref{eq7} and \eqref{eq8}
respectively. Considering the definitions \eqref{eq6} and
\eqref{eq18} under the adjoint representation \eqref{eq68} of $s$
which we have proved to exist in the previous section we can
immediately see the identification
\begin{equation}\label{eq76}
e^{\mathbf{\Gamma }}e^{\mathbf{\Lambda }}=e^{\frac{1}{2}\phi
^{i}\widetilde{g}_{i}}e^{\chi ^{\alpha}\widetilde{f}_{\alpha}}
\equiv e^{\frac{1}{2}\phi ^{i}H_{i}}e^{\chi
^{\alpha}E_{\alpha}}=\nu,
\end{equation}
where $\nu$ is the coset representative of the sigma model
\cite{nej1,nej2}. Thus when the representation defined in
\eqref{eq68} is chosen which sends
\begin{equation}\label{eq77}
H_{i}\longrightarrow \widetilde{g}_{i}\quad ,\quad
E_{\alpha}\longrightarrow \widetilde{f}_{\alpha},
\end{equation}
the set of first-order equations \eqref{eq73} can be written as
\begin{equation}\label{eq78}
\ast \overset{\rightharpoonup }{\mathbf{\Psi
}}=(-1)^{D}\nu\overset{\rightharpoonup }{\mathbf{A}}.
\end{equation}
Now if we take the exterior derivative of both sides we get
\begin{equation}\label{eq79}
d(\ast \overset{\rightharpoonup }{\mathbf{\Psi
}})=(-1)^{D}d\nu\overset{\rightharpoonup }{\mathbf{A}},
\end{equation}
where we have used $d\mathbf{A}^{m}=0$. Since \eqref{eq78} is a
vector equation it can be written as\footnote{Note that $\nu^{-1}$
exists by definition.}
\begin{equation}\label{eq80}
\overset{\rightharpoonup }{\mathbf{A}}=(-1)^{D}\nu^{-1}\ast
\overset{\rightharpoonup }{\mathbf{\Psi }}.
\end{equation}
Inserting this into \eqref{eq79} we get
\begin{equation}\label{eq81}
d(\ast \overset{\rightharpoonup }{\mathbf{\Psi
}})=d\nu\nu^{-1}\ast \overset{\rightharpoonup }{\mathbf{\Psi }}.
\end{equation}
In this equation we readily realize that
\begin{equation}\label{eq82}
\mathcal{G}=d\nu\nu^{-1},
\end{equation}
is the Cartan-form induced by the coset map $\nu$ and it is
explicitly calculated in \cite{nej2}. It reads
\begin{equation}\label{eq83}
\mathcal{G}=\frac{1}{2}d\phi ^{i}H_{i}+e^{%
\frac{1}{2}\beta _{i}\phi
^{i}}\mathbf{\Omega}^{\beta}_{\alpha}d\chi^{\alpha}E_{\beta },
\end{equation}
where
\begin{equation}\label{eq84}
\begin{aligned}
 \mathbf{\Omega}&=\sum\limits_{m=0}^{\infty }\dfrac{\omega
^{m}}{(m+1)!}\\
\\
&=(e^{\omega}-I)\,\omega^{-1}.
\end{aligned}
\end{equation}
Here the (S-r)$\times$(S-r) matrix $\omega$ has the components
\begin{equation}\label{eq85}
\omega_{\beta}^{\gamma}=\chi^{\alpha}K_{\alpha\beta}^{\gamma},
\end{equation}
with $K_{\alpha\beta}^{\gamma}$ defined as
\begin{equation}\label{eq86}
[E_{\alpha},E_{\beta}]=K_{\alpha\beta}^{\gamma}E_{\gamma}.
\end{equation}
Since we chose the representation generated by \eqref{eq69} we can
write \eqref{eq83} as
\begin{equation}\label{eq87}
\mathcal{G}=\frac{1}{2}d\phi ^{i}\widetilde{g}_{i}+e^{%
\frac{1}{2}\beta _{i}\phi
^{i}}\mathbf{\Omega}^{\beta}_{\alpha}d\chi^{\alpha}\widetilde{f}_{\beta}.
\end{equation}
If we insert this back in \eqref{eq81} we obtain (in component
form)
\begin{equation}\label{eq88}
d(\ast \mathbf{\Psi
}^{m})=(\frac{1}{2}d\phi ^{i}\widetilde{g}_{in}^{m}+e^{%
\frac{1}{2}\beta _{i}\phi
^{i}}\mathbf{\Omega}^{\beta}_{\alpha}d\chi^{\alpha}\widetilde{f}_{\beta
n}^{m})\wedge\ast\mathbf{\Psi }^{n}.
\end{equation}
For $1\leq m\leq r$ \eqref{eq88} yields
\begin{equation}\label{eq89}
\frac{1}{2}d(\ast d\phi^{i})=\frac{1}{2}d\phi ^{j}\wedge\widetilde{g}_{jn}^{i}\ast\mathbf{\Psi }^{n}+e^{%
\frac{1}{2}\beta _{k}\phi
^{k}}\mathbf{\Omega}^{\beta}_{\alpha}d\chi^{\alpha}\wedge\widetilde{f}_{\beta
n}^{i}\ast\mathbf{\Psi }^{n}.
\end{equation}
From \eqref{eq8} the first term on the RHS vanishes and if we
split the sum in the second term on the index $n$ then we have
\begin{equation}\label{eq90}
\frac{1}{2}d(\ast d\phi^{i})=e^{%
\frac{1}{2}\beta _{k}\phi
^{k}}\mathbf{\Omega}^{\beta}_{\alpha}d\chi^{\alpha}\wedge(\widetilde{f}_{\beta
j}^{i}\ast\mathbf{\Psi }^{j}+\widetilde{f}_{\beta,\gamma+r
}^{i}\ast\mathbf{\Psi }^{\gamma+r}).
\end{equation}
Now from \eqref{eq7} again the first term on the RHS vanishes.
Further index splitting in the sum on the remaining term gives
\begin{equation}\label{eq91}
\frac{1}{2}d(\ast d\phi^{i})=e^{%
\frac{1}{2}\beta _{k}\phi
^{k}}\mathbf{\Omega}^{\beta}_{\alpha}d\chi^{\alpha}\wedge(\widetilde{f}_{\beta,\beta+r
}^{i}\ast\mathbf{\Psi
}^{\beta+r}+\underset{\kappa\neq\beta}{\sum}\widetilde{f}_{\beta,\kappa+r
}^{i}\ast\mathbf{\Psi }^{\kappa+r}).
\end{equation}
Due to \eqref{eq7} the second sum on the RHS also vanishes. By
reading\footnote{The reader should pay attention that the first
term inside the parentheses on the RHS of \eqref{eq91} is not a
sum but a single term.} the value of $\widetilde{f}_{\beta,\beta+r
}^{i}$ from \eqref{eq7} and also by using \eqref{eq74} we finally
get
\begin{equation}\label{eq92}
d(\ast d\phi^{i})=\frac{1}{2}\underset{\alpha,\beta,\gamma\in \Delta_{nc}^{+}}{\sum}\beta_{i}e^{%
\frac{1}{2}\beta _{k}\phi
^{k}}\mathbf{\Omega}^{\beta}_{\alpha}d\chi^{\alpha}\wedge e^{%
\frac{1}{2}\beta _{j}\phi
^{j}}\mathbf{\Omega}^{\beta}_{\gamma}\ast d\chi^{\gamma},
\end{equation}
where $i$ is the free index. The set of equations in \eqref{eq92}
are exactly the second-order dilaton field equations of the sigma
model which are derived in \cite{nej1,nej2,ker1,ker2}. Now on the
other hand if we consider \eqref{eq88} for $m>r$ then we have
\begin{equation}\label{eq93}
d\ast(e^{\frac{1}{2}\beta _{i}\phi
^{i}}\mathbf{\Omega}^{\beta}_{\alpha}d\chi^{\alpha} )=(\frac{1}{2}d\phi ^{i}\widetilde{g}_{in}^{\beta+r}+e^{%
\frac{1}{2}\kappa_{i}\phi
^{i}}\mathbf{\Omega}^{\kappa}_{\alpha}d\chi^{\alpha}\widetilde{f}_{\kappa
n}^{\beta+r})\wedge\ast\mathbf{\Psi }^{n}.
\end{equation}
In this equation $\beta$ is the free index. Similar to our
calculation above again due to \eqref{eq7} and \eqref{eq8} after
eliminating the vanishing terms on the RHS we get
\begin{equation}\label{eq94}
d\ast(e^{\frac{1}{2}\beta _{i}\phi
^{i}}\mathbf{\Omega}^{\beta}_{\alpha}d\chi^{\alpha}
)=-\frac{1}{2}\beta_{i}d\phi ^{i}\wedge\ast\mathbf{\Psi
}^{\beta+r}+e^{\frac{1}{2}\kappa_{i}\phi
^{i}}\mathbf{\Omega}^{\kappa}_{\alpha}d\chi^{\alpha}\widetilde{f}_{\kappa,\gamma+r}^{\beta+r}\wedge\ast\mathbf{\Psi
}^{\gamma+r}.
\end{equation}
Furthermore by using \eqref{eq74} and \eqref{eq7} we finally have
\begin{equation}\label{eq95}
\begin{aligned}
d\ast(e^{\frac{1}{2}\beta _{i}\phi
^{i}}\mathbf{\Omega}^{\beta}_{\alpha}d\chi^{\alpha}
)&=-\frac{1}{2}\beta_{i}d\phi
^{i}\wedge(e^{\frac{1}{2}\beta_{j}\phi
^{j}}\mathbf{\Omega}^{\beta}_{\alpha}\ast
d\chi^{\alpha})\\
\\
&\:\:\:\:\:+\underset{\kappa-\theta=-\beta}{\sum}e^{\frac{1}{2}\kappa_{i}\phi
^{i}}\mathbf{\Omega}^{\kappa}_{\alpha}d\chi^{\alpha}\wedge
(N_{\kappa,-\theta}e^{\frac{1}{2}\theta_{j}\phi
^{j}}\mathbf{\Omega}^{\theta}_{\sigma}\ast d\chi^{\sigma}),
\end{aligned}
\end{equation}
where on the RHS in the second term the sum is on the index
$\kappa\in\Delta_{nc}^{+}$ (again $\beta$ is the free index) and
due to \eqref{eq10} the index $\theta\in\Delta_{nc}^{+}$ (if the
corresponding root exists in $\Delta_{nc}^{+}$ when one fixes
$\beta$ and $\kappa$) must be chosen according to the root
condition stated above\footnote{Of course from \eqref{eq10} if
such a root $\theta$ does not exist in $\Delta_{nc}^{+}$ then that
term for a particular choice of $\beta$ and $\kappa$ is zero. We
should also state that there is no sum on the index $\theta$ in
\eqref{eq95} instead if it exists $\theta$ becomes fixed when
$\beta$ and $\kappa$ are chosen.}. As the index (the root) $\beta$
runs in $\Delta_{nc}^{+}$ these equations are the second-order
axion field equations of the sigma model which are derived in
\cite{nej1,nej2,ker1,ker2}.
\section{Conclusion}
We have presented a rigorous proof which denotes that the dualized
algebra of the coset sigma model with globally Riemannian
symmetric target space is indeed a Lie algebra. Although the
commutation relations of the dualized coset algebra were derived
in \cite{nej2} their Lie algebra structure was not proved in that
work. By showing that the structure constants or the commutators
indeed satisfy the Jacobi identities we have justified the Lie
algebra notion of the dualized algebra which is an extension of
the original coset algebra of the sigma model. Later we have also
mentioned that the dualized coset algebra which contains the
ordinary one in it and therefore which can be considered as a
deformation admits an adjoint representation for the original
coset algebra. Finally under this special representation we have
shown that the second-order field equations can be obtained by
differentiating the first-order equations which appeared in
\cite{nej1,nej2} as consistency conditions of the dualization
construction. Therefore we have proved that these consistency
conditions are the true algebraic first-order equations of the
corresponding sigma model.

As it can be inferred from the sequence of its sections this paper
aims to show that under a special representation generated by a
duality algebra the second-order Euler-Lagrange equations can be
integrated to obtain first-order field equations. These
first-order equations are already derived (in other words
suggested) in \cite{nej1,nej2}. However only via this work they
are proven to be the algebraically correct ones since in
\cite{nej1,nej2} they appeared as a consistency condition embedded
within the dualization of the theory. On the other hand in this
work we have proven that if one applies an exterior derivative on
these first-order equations one gets the correct second-order
field equations which are the Euler-Lagrange ones. Depending on
the analysis given here we can easily state that the dualization
of a theory apart from its enlarged geometrical construction is an
efficient way of inventing the correct representation of the coset
algebra so that this representation leads to an integration of the
field equations. In other words within the dualized theory the
original coset algebra is implemented in a Lie algebra deformation
of it (the dualized coset algebra) in such a way that the
generated adjoint representation becomes an appropriate one in
which the integration of the field equations exists. We have shown
that the representation which enables to construct the first-order
field equations depends on the Lie algebra structure of the
dualized coset algebra  which is a special and a non-trivial
extension of the original one. In \cite{nej1,nej2} this algebra
and the first-order field equations were derived from a partially
geometrical point of view. Here by showing the legacy of the Lie
algebra structure and accordingly the adjoint representation we
have algebraically complemented the achievements of
\cite{nej1,nej2}. Thus our exact proof additionally justifies the
correctness of the first-order field equations of the symmetric
space sigma model. This is an essential result as it enables the
reduction of order of the second-order partial differential field
equations of the sigma model which is an important ingredient in
supergravity as well as string theory also in QFT.

From the analysis point of view another essential result of the
present work can be considered as the generation of an extended
Lie algebra structure starting form a solvable Lie subalgebra of a
Lie algebra. In fact the arguments of section two can easily be
generalized and depending on the complete and the formal proof
presented in section two we can state that every subalgebra of the
Borel subalgebra of a Lie algebra sits in another Lie algebra with
a doubled dimension.

\end{document}